\documentstyle[prl,preprint,aps,epsf]{revtex}
\begin{document}

\title{Universality or Non--Universality of Absorption Cross Sections 
      for 
Extended Objects}
\author{D.K. Park$^{1,2}$\footnote{e-mail:
dkpark@hep.kyungnam.ac.kr}
and
H. J. W. M\"{u}ller-Kirsten$^{1}$\footnote{e-mail:
mueller1@physik.uni-kl.de}}
\address{1. Department of Physics,
 University of Kaiserslautern, D-67653 Kaiserslautern, Germany\\
2.Department of Physics, Kyungnam University, Masan, 631-701, Korea}

\maketitle

\begin{abstract}
The calculation of absorption cross sections for minimal scalars in
supergravity backgrounds is an important aspect of the investigation of
AdS/CFT correspondence and requires a matching of appropriate wave 
functions. The low energy case has attracted particular attention.
 In the following 
the dependence of the cross section on the matching point is investigated.
It is shown that the low energy limit is independent of the matching point
and hence exhibits universality. In the high energy limit the independence 
is not maintained, but the result is believed to possess the correct energy 
dependence. 
\end{abstract}

\vspace{1cm}

\centerline{PACS Numbers:04.70.Dy, 04.62.+v, 11.25.-w}

\newpage

\vspace{3cm}
After the entropy problem was solved within the framework of string 
theory\cite{stro96} by identifying extremal black holes with BPS states, 
recent interest seems to be shifted to the Hawking radiation 
problem\cite{hawk75}. In this context the absorption cross section of extended 
objects has been computed in the framework of various 
models\cite{das96-1,das96-2,gub96,kle97,cve00,man00,park00} requiring matching
of wave functions, and the result always coincides with the area of the 
horizon up to a constant in the low energy limit. This universality\footnote{
While the conventional meaning of universality
 indicates that the low energy cross
section coincides with the area of the horizon, we will use this terminology
when the low energy cross section exhibits a common behavior.} is 
examined in general for a spherically symmetric and asymptotically flat
geometry\cite{das97}, and is in addition generalized by computing the 
frequency--dependent leading order\cite{emp98}.

In this letter we argue that this universality property at low energy is 
related to the insensitivity of extended objects to the matching equations
between the asymptotic solution $\phi_{\omega}^{\infty}$ and the 
near-horizon solution $\phi_{\omega}^{near}$. Although this universal
property disappears in the high enery limit, it will be shown that 
even in this case one can obtain the important information, {\it i.e.}
the explicit energy-dependence of the absorption rate. Also, it is 
briefly shown that the universality
 property is maintained for the massive scalar 
case also. 

We consider a massless scalar field $\Phi$ minimally coupled to a spherically 
symmetric geometry
\begin{equation}
\label{geometry}
ds^2 = \gamma_{\mu \nu}(r) d x^{\mu} d x^{\nu} + f(r) d r^2 + r^2 h(r)
d \Omega_{n+1}
\end{equation}
where $\gamma_{\mu \nu}(r) (\mu, \nu = 0, \cdots, p)$ is the metric on a 
$(p+1)$--dimensional world volume of the extended objects. The geometry
is assumed to be asymptotically flat: $\gamma_{\mu \nu}(r) \rightarrow
\eta_{\mu \nu}$, $f(r), g(r) \rightarrow 1$ as $r \rightarrow \infty$.
Introducing a tortoise coordinate $r^{\ast}$ defined by 
$d r^{\ast} \equiv d r \sqrt{- \gamma^{t t}(r) f(r)}$, and considering only
$s$-waves, {\it i.e.} $\Phi = e^{-i \omega t} \phi_{\omega}(r)$,
 one can derive a differential equation similar to the Schr\"odinger equation
\begin{equation}
\label{schrodinger}
\left[ - \frac{d^2}{d r^{\ast 2}} + V \right] \psi = \omega^2 \psi
\end{equation}
where
\begin{eqnarray}
\label{exp1}
\phi_{\omega}&=& \frac{1}{\sqrt{U}} \psi,    \\   \nonumber
U&=& \left[\gamma \gamma^{t t} \left[r^2 h(r) \right]^{n+1} \right]^{\frac{1}{2}},    \\   \nonumber
V&=& \frac{1}{\sqrt{U}} \frac{d^2 \sqrt{U}}{d r^{\ast 2}}
\stackrel{r^{\ast}\rightarrow\infty}{\longrightarrow}\frac{n^2-1}{4r^2}.
\end{eqnarray}
Here, $\gamma \equiv det \gamma_{\mu \nu}$. The solution at $r \rightarrow
\infty$ is easily obtained in terms of Bessel functions,
\begin{equation}
\label{asysol}
\phi_{\omega}^{\infty} = \frac{1}{(\omega r)^{n/2}}
\left[ A J_{\frac{n}{2}} (\omega r) + B J_{-\frac{n}{2}} (\omega r) \right]
\end{equation}
for odd $n$. Of course, for even $n$ the Bessel function with negative order 
has to be replaced by the Neumann function. Since the final forms of the
absorption cross section are always equivalent, we will consider only
the $n = odd$ case. Using the asymptotic formula of the Bessel function it is
straightforward to derive incoming and outgoing fluxes;
\begin{eqnarray}
\label{asyflux}
{\cal F}_{\infty}^{in}&=& - \frac{1}{\pi \omega^n}
\left[ \mid A \mid^2 + \mid B \mid^2 + A B^{\ast} e^{i \frac{n}{2} \pi}
+ A^{\ast} B e^{-i \frac{n}{2} \pi} \right]  \\  \nonumber
{\cal F}_{\infty}^{out}&=&  \frac{1}{\pi \omega^n}
\left[ \mid A \mid^2 + \mid B \mid^2 + A^{\ast} B e^{i \frac{n}{2} \pi}
+ A B^{\ast} e^{-i \frac{n}{2} \pi} \right].
\end{eqnarray}

In order to obtain the near--horizon solution we introduce several 
parameters as in Ref.\cite{emp98}:
\begin{eqnarray}
\label{parameters}
\lim_{r \rightarrow 0} U(r) &\approx& S r^{a - b}   \\  \nonumber
\lim_{r \rightarrow 0} \sqrt{- \gamma^{t t}(r) f(r)} &\approx&
\frac{T}{r^{b+1}}
\end{eqnarray}
and we confine ourselves to the case of $0 < b \leq a$\cite{emp98}.
Then, it is straightforward to derive a near--horizon solution in terms of a
Hankel function,
\begin{equation}
\label{near}
\phi_{\omega}^{near} \approx 
\frac{1}{(\omega r)^{\frac{a}{2}}} H_{\frac{a}{2 b}}^{(2)} 
\left( \frac{\omega T}{b r^b} \right)
\end{equation}
and the incoming flux is 
\begin{equation}
\label{influx}
{\cal F}^{near} = \frac{4 b S}{\pi \omega^a T}.
\end{equation}
In deriving Eq.(\ref{near}) we used the boundary condition that, as $r$ 
approaches zero, the field contains only incoming waves. Since the absorption
cross section per unit volume is defined as 
\begin{equation}
\label{defsection}
\sigma = \frac{(2 \pi)^{n+1}}{\omega^{n+1} \Omega_{n+1}}
         \mid \frac{{\cal F}^{near}}{{\cal F}_{\infty}^{in}} \mid
\end{equation}
where $\Omega_{n+1} = 2 \pi^{1 + \frac{n}{2}} / \Gamma(1 + \frac{n}{2})$, it 
is completely determined by determining the coefficients $A$ and $B$ through a 
matching equation. As claimed above we will show that the absorption 
cross section in the low energy limit is very insensitive to the choice of 
matching equation, which results in the universality of the low--energy 
absorption cross section. 

To show this we assume there is a matching point in the finite region, say
at $r = R$ and use the usual quantum mechanical matching method, 
{\it i.e.} continuity of the wave function and its derivative;
\begin{eqnarray}
\label{matching1}
\phi_{\omega}^{\infty}(R)&=&\phi_{\omega}^{near}(R)  \\   \nonumber
\frac{d}{dR} \phi_{\omega}^{\infty}(R)&=& \frac{d}{dR}
\phi_{\omega}^{near}(R).
\end{eqnarray}
Then it is straightforward to obtain $A$ and $B$:
\begin{eqnarray}
\label{coef}
A&=& (-1)^{\frac{n - 1}{2}}
    \frac{\pi R (\omega R)^{\frac{n - a}{2}}}{2}
    \Bigg[ \frac{n - a}{2 R} J_{-\frac{n}{2}} (\omega R)
           H_{\frac{a}{2 b}}^{(2)} \left( \frac{\omega T}{b R^b} \right)
          - \omega J_{-\frac{n}{2}}^{\prime} (\omega R)
            H_{\frac{a}{2 b}}^{(2)} \left( \frac{\omega T}{b R^b} \right)
                                                             \\  \nonumber
& &\hspace{6.0cm}     - \frac{\omega T}{R^{b + 1}} J_{-\frac{n}{2}} (\omega R)
            H_{\frac{a}{2 b}}^{(2) \prime} \left( \frac{\omega T}{b R^b} \right)
                                                \Bigg]
                                                      \\   \nonumber
B&=& (-1)^{\frac{n + 1}{2}}
    \frac{\pi R (\omega R)^{\frac{n - a}{2}}}{2}
    \Bigg[ \frac{n - a}{2 R} J_{\frac{n}{2}} (\omega R)
           H_{\frac{a}{2 b}}^{(2)} \left( \frac{\omega T}{b R^b} \right)
          - \omega J_{\frac{n}{2}}^{\prime} (\omega R)
            H_{\frac{a}{2 b}}^{(2)} \left( \frac{\omega T}{b R^b} \right)
                                                          \\   \nonumber
& &\hspace{6.0cm}     - \frac{\omega T}{R^{b + 1}} J_{\frac{n}{2}} (\omega R)
            H_{\frac{a}{2 b}}^{(2) \prime} \left( \frac{\omega T}{b R^b} \right)
                                                \Bigg]
\end{eqnarray}
where a prime denotes differentiation with respect to the argument. 
Using Eq.(\ref{coef}) one can plot the cross section for 
different values of $R$ as shown in Fig. 1. Fig. 1 demonstrates the important
fact that the low energy cross section is independent of
the choice of the
matching point $R$, which is the origin of the idea of universality. 

To show this more explicitly we compute the coefficients $A$ and $B$ in the 
low energy limit by using the asymptotic formulas of Bessel and Hankel
functions:
\begin{eqnarray}
\label{lowcoef}
A&=&(-1)^{\frac{n-1}{2}} i n \omega^{-\frac{a}{2}} 2^{\frac{n}{2} - 1}
\left( \frac{2 b}{\omega T}\right)^{\frac{a}{2 b}}
\frac{\Gamma \left( \frac{a}{2 b} \right)}{\Gamma \left( 1 - \frac{n}{2}
                                                    \right)}
                                                     \\   \nonumber
B&=&0
\end{eqnarray}
at the leading order. One should note that the $R$-dependence disappears in 
$A$ and $B$. Computing the low energy cross section using Eq.(\ref{lowcoef}),
one can obtain straightforwardly
\begin{equation}
\label{lowsec}
\sigma_{L} = \frac{\pi}{\Gamma^2 \left( \frac{a}{2 b} \right)} 
S \Omega_{n+1} \left( \frac{\omega T}{2 b} \right)^{\frac{a}{b} - 1}
\end{equation}
which coincides with the result of Ref.\cite{emp98}. Of course, when 
$a = b$, $\sigma_L$ becomes $S \Omega_{n+1}$ which is an area of horizon.

One may argue that this $R$-dependence of $\sigma_L$ is a special 
property of the matching equation (\ref{matching1}). To disprove this one may 
choose other matching equations such as 
\begin{eqnarray}
\label{matching2}
\mid \frac{{\cal F}_{\infty}^{out}}{{\cal F}_{\infty}^{in}} \mid
  &+& \mid \frac{{\cal F}^{near}}{{\cal F}_{\infty}^{in}} \mid = 1
                                                    \\   \nonumber
\phi_{\omega}^{\infty} (R)&=& \phi_{\omega}^{near} (R).
\end{eqnarray}
However, after tedious calculation one can show that this matching 
equation also leads to Eq. (\ref{lowsec}) in the low energy limit.
We think this insensitivity of extended objects to the choice of matching 
equations and matching points results in the universality of this limit.

However, the situation is completely different in the high energy limit.
Taking the high energy limit,{\it i.e.} $\omega \rightarrow \infty$, in
Eq. (\ref{coef}), one can obtain
\begin{eqnarray}
\label{highcoef}
A&=& (-1)^{\frac{n-1}{2}} (\omega R)^{\frac{n-a+1}{2}} 
\left( \frac{b R^b}{\omega T} \right)^{\frac{1}{2}}
e^{-i \left[ \frac{\omega T}{b R^b} - \frac{\pi}{4} \left( 1 + \frac{a}{b}
                                                     \right) \right] }
                                                      \\   \nonumber
& & \hspace{2cm} \times
\left[\sin \left( \omega R + \frac{n-1}{4} \pi \right) + 
      \frac{i T}{R^{b+1}} \cos \left( \omega R + \frac{n-1}{4} \pi \right)
                                                          \right]
                                                      \\   \nonumber
B&=& (-1)^{\frac{n+1}{2}} (\omega R)^{\frac{n-a+1}{2}}
\left( \frac{b R^b}{\omega T} \right)^{\frac{1}{2}}
e^{-i \left[ \frac{\omega T}{b R^b} - \frac{\pi}{4} \left( 1 + \frac{a}{b}
                                                     \right) \right] }
                                                      \\   \nonumber
& & \hspace{2cm} \times
\left[\sin \left( \omega R - \frac{n+1}{4} \pi \right) +
      \frac{i T}{R^{b+1}} \cos \left( \omega R - \frac{n+1}{4} \pi \right)
                                                          \right],
\end{eqnarray}
which yields the absorption cross section $\sigma_H$ in the high energy
limit to be
\begin{equation}
\label{highsec}
\sigma_H = 
\frac{(2 \pi / \omega)^{n+1}}{\Omega_{n+1} R^{n-a}}
\frac{4 S / T}{\left[ \sqrt{ \frac{R^{b+1}}{T}} - 
                      \sqrt{\frac{T}{R^{b+1}}} \right]^2}.
\end{equation}
The appearance of $R$ in Eq. (\ref{highsec}) indicates the high energy cross
section loses the universality property. However, the $\omega$-dependence of 
$\sigma_H$, {\it i.e.} $\sigma_H \propto \omega^{-(n+1)}$, exhibits a 
decreasing behavior. This decreasing behavior in the high energy limit 
is also found numerically in Ref.\cite{cve00}. This can be an important
property as we learned from blackbody radiation. One may question the 
credibility of this $\omega$-dependence in view of the $R$-dependence of 
$\sigma_H$. In fact, this is our belief, and the rigorous proof is still
an open problem. However, one can achieve some more credibility by considering
more complicated situations such as a fixed scalar whose low energy 
cross section does not exhibit a universality\cite{kol97}. The authors of 
Ref.\cite{kol97} compute the low energy absorption cross section by matching
$\phi_{\omega}^{near}$ and $\phi_{\omega}^{\infty}$ through the solution
in the intermediate region as Unruh did in his seminal paper\cite{unruh76}
and obtained $\sigma_s = 2 \pi \omega^2$ for the $s$-wave. If one applies our
matching method to this problem, $\sigma = 2 \pi \omega^2 R^2 / (R-1)^2$ is 
obtained. Although the explicit dependence on $R$ in $\sigma$ indicates the
non-universality in this case, apart from this $R$-dependent factor the 
cross section exhibits the correct $\omega$-dependence. This is the reason
why we  can have confidence in the $\omega$-dependence of $\sigma_H$ in
Eq. (\ref{highsec}).

Finally, we comment on the absorption cross section for the case of
a massive 
scalar. It is interesting to know whether the universal property of 
the low energy cross section is still maintained or not. In this case
the potential in Eq. (\ref{exp1}) is changed to 
\begin{equation}
\label{masspoten}
V = \frac{1}{\sqrt{U}} \frac{d^2 \sqrt{U}}{d r^{\ast 2}} - \frac{m^2}
                                                           {\gamma^{t t}}
\end{equation}
where $m$ is the mass of the scalar field. The asymptotic solution
in this case is the same as that of Eq. (\ref{asysol}) if $\omega r$ is
replaced by $\omega v r$ where $v = \sqrt{1 - m^2 / \omega^2}$.

If we assume $\lim_{r \rightarrow 0} \gamma^{t t}(r) \approx - W / r^{2 c}$
where $W$ and $c$ are some constants, the potential of Eq. (\ref{schrodinger})
in the $r \rightarrow 0$ region becomes of the form
\begin{equation}
\label{potentwo}
V = V_1(r) + V_2(r)
\end{equation}
where
\begin{eqnarray}
\label{v1v2}
V_1(r)&=& \frac{a^2 - b^2}{4 T^2} r^{2 b}
                                            \\   \nonumber
V_2(r)&=& \frac{m^2}{W} r^{2c}.
\end{eqnarray}
We consider only the $b < c$ case for simplicity. The full description
of the massive scalar case will be discussed elsewhere. Then we can take
$V \approx V_1$ approximately, and hence the near-horizon solution is 
unchanged. This means the mass effect is decoupled in this case in the
$r \approx 0$ region. 

By applying our method it is straightforward to obtain the low energy
cross section $\sigma_{L}^{m} = v^n \sigma_L$. Of course, $\sigma_L^m$ 
becomes $v^n$ times the area of the horizon when $a = b$. Also, in the 
high energy limit we can obtain the same cross section as that of 
Eq. (\ref{highsec}) if $R^{b+1}$ in the square root is replaced by 
$v R^{b+1}$.

In conclusion we make the following remarks.
The explicit and exact calculation of S--matrices for
specific potentials is generally only possible in some
special cases and requires
a detailed study of the solutions of the appropriate
wave equation in adjoining domains of validity over the
entire range of the variable. This old problem
which in the past was studied in $1+3$ dimensions
has received fresh impetus from the string
theory interest in absorption cross
sections and also for other and higher dimensions. In the
special case of the $D3$ brane the absorption cross section
can be calculated explicitly in terms of modified
Mathieu functions in both the low and the high energy
domains\cite{man00,park00}. The matching of different branches
of the solutions in domains of
overlap can be done but is nontrivial.  It is natural,
therefore, particularly if one is interested, for instance, only in the
low energy case, to devise simpler methods for
the derivation.  The method of using Bessel and Hankel functions
is such a method, and has been employed particularly frequently
in this context for asymptotically flat metrics\cite{kle97,kleb}.
However, since the matching point can be chosen
arbitrarily, one wants to be
ascertained that the result does not depend 
on its choice.  In the above we demonstrated
for a wide class of metrics the universality of
the low energy result, i.e. its
independence of the matching point.
Applying the method to the high energy domain we see
that there this universality does not ensue, although
the energy dependence is expected to be
correct.  In a further extension, the method
has also been applied above to massive scalars. Our
findings are therefore particularly reensuring for
the application of the simple Bessel function method
to the low energy case. In view of the possible
wide applicability of the method, the demonstration
of universality is also of general interest.

{\bf Acknowledgement:} D.K. P. acknowledges support by the
Deutsche Forschungsgemeinschaft (DFG).

\newpage
\centerline{\bf Figure Caption:}
\vspace{0.4cm}
\noindent
{\bf Fig. 1}

\noindent
The absorption cross section for $n=1, b=T=s=1$
and $a=1$ and $2$ for $R=1,2,3,4$.

\newpage
\epsfysize=5cm \epsfbox{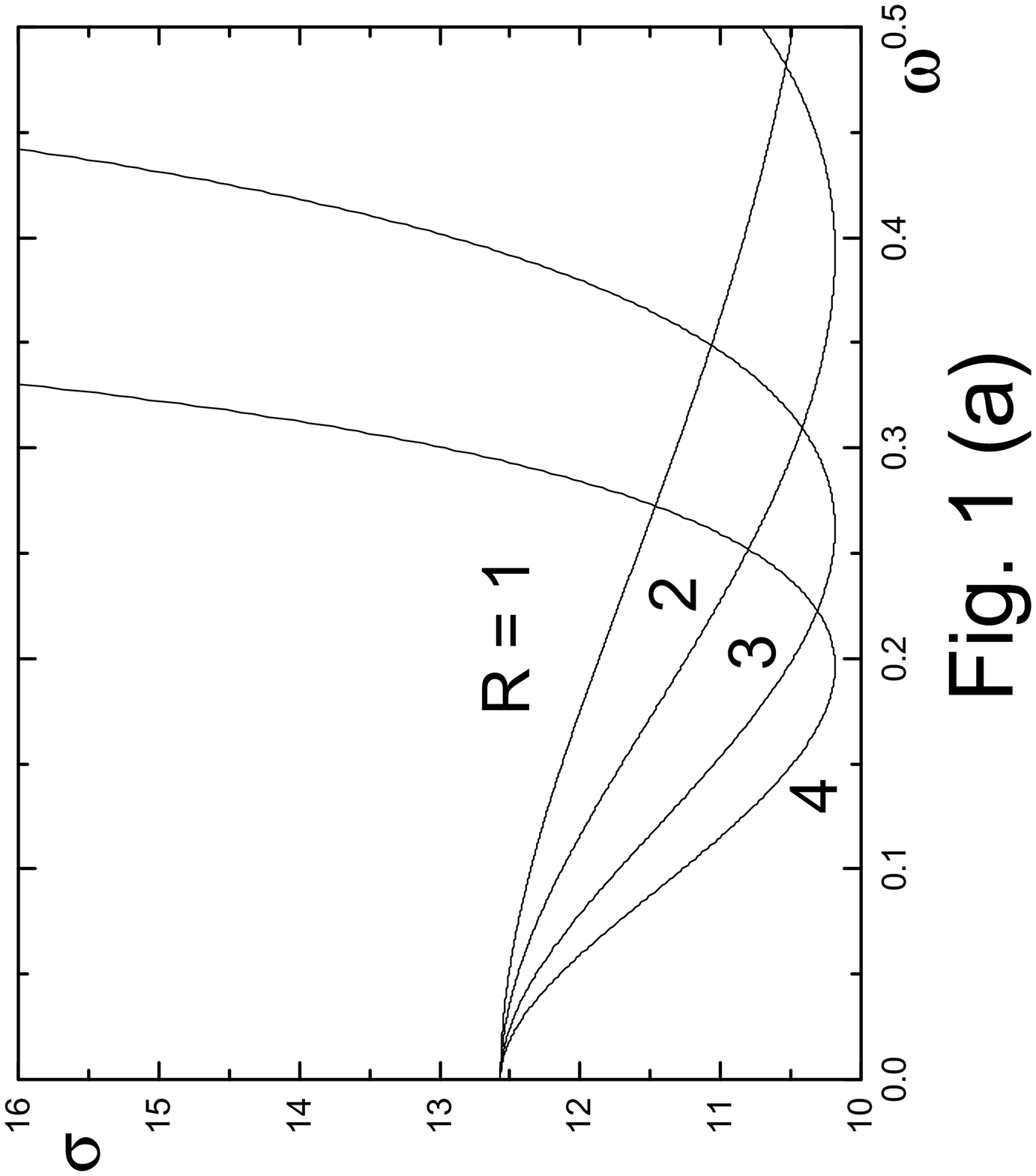}
\newpage 
\epsfysize=5cm \epsfbox{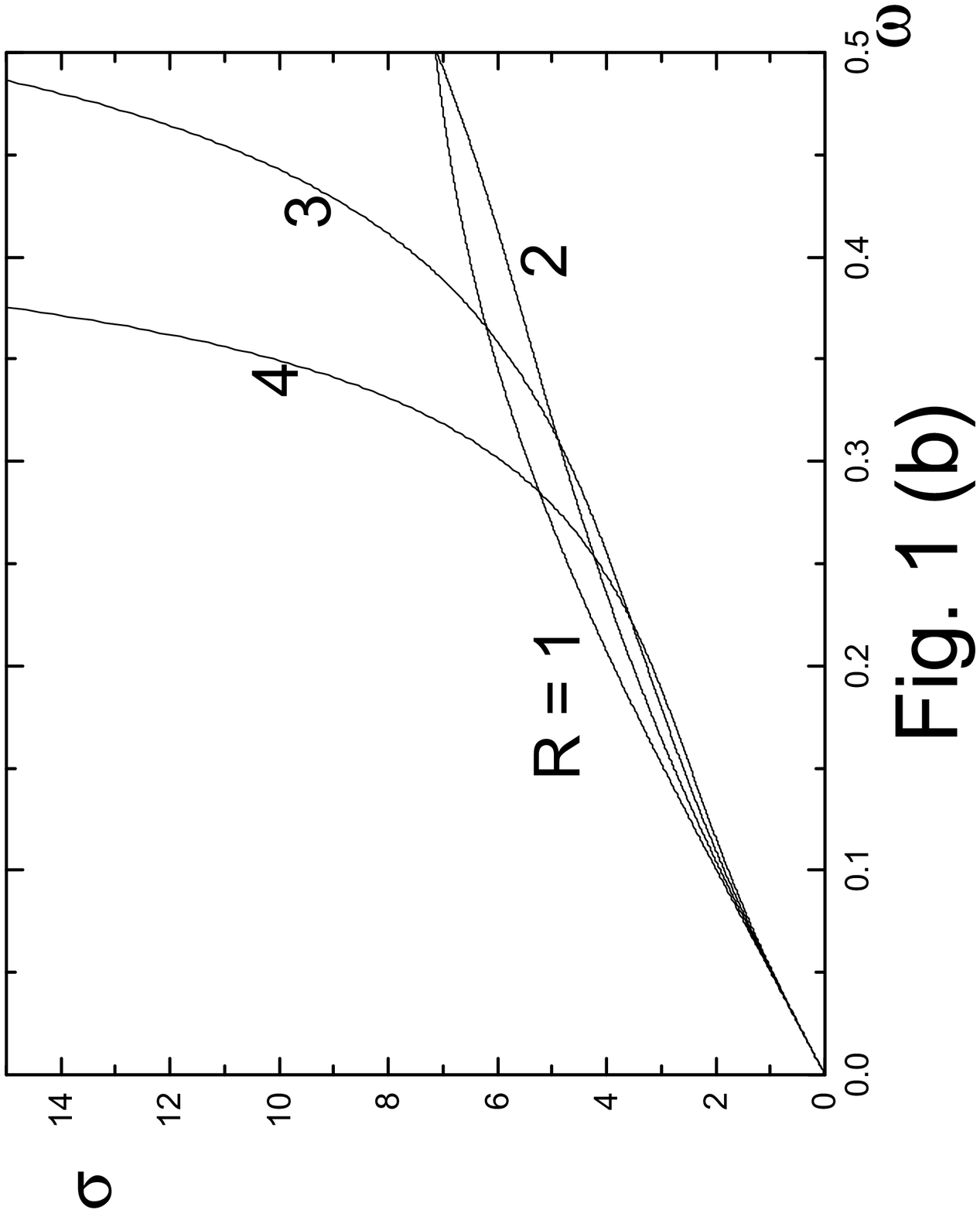}

\begin{thebibliography}{99}
\bibitem{stro96} A. Strominger and C. Vafa, Phys. Lett. {\bf B379}, 99 (1996).
\bibitem{hawk75} Commun. Math. Phys. {\bf 43}, 199 (1975). 
\bibitem{das96-1} S. R. Das and S. D. Mathur, hep-th/9606185, Nucl. Phys. 
{\bf B478}, 561 (1996).
\bibitem{das96-2} S. R. Das and S. D. Mathur, hep-th/9607149, Nucl. Phys. 
{\bf B482}, 153 (1996). 
\bibitem{gub96} S. S. Gubser and I. R. Klebanov, hep-th/9608108, Nucl. Phys. 
{\bf B482}, 173 (1996).
\bibitem{kle97} I. R. Klebanov, hep-th/9702076, Nucl. Phys. 
{\bf B496}, 231 (1997).  
\bibitem{cve00} M. Cvetic, H. L\"u, and J. F. Vazquez-Poritz, hep-th/0002128. 
\bibitem{man00} R. Manvelyan, H. J. W. M\"{u}ller-Kirsten, J. -Q. Liang, and
Y. Zhang, hep-th/0001179, Nucl. Phys. {\bf B579}, 177 (2000).
\bibitem{park00} D. K. Park, S. N. Tamaryan, H. J. W. M\"{u}ller-Kirsten, and
J. Zhang, hep-th/0005165.
\bibitem{das97} S. R. Das, G. Gibbons and S. D. Mathur, hep-th/9609052, Phys. 
Rev. Lett. {\bf 78}, 417 (1997).
\bibitem{emp98} R. Emparan, hep-th/9706204, Nucl. Phys. {\bf B516}, 297 (1998).
\bibitem{kol97} B. Kol and R. Rajaraman, hep-th/9608126, Phys. Rev. {\bf D56},
983 (1997).
\bibitem{unruh76} W. G. Unruh, Phys. Rev. {\bf D14}, 3251 (1976).
\bibitem{kleb} I.R. Klebnaov, W. Taylor IV and M.Van
 Raamsdonk, hep--th/9905174, Nucl. Phys. {\bf B560}, 207 (1999).
\end{thebibliography}
\end{document}